\documentstyle[epsfig]{elsart}

\newcommand \be{\begin{equation}}
\newcommand \ba{\begin{eqnarray}}
\newcommand \ee{\end{equation}}
\newcommand \ea{\end{eqnarray}}

\begin{document}
\runauthor{Zhou and Sornette}
\begin{frontmatter}
\title{2000-2003 Real Estate Bubble in the UK but not in the USA}
\author[iggp]{\small{Wei-Xing Zhou}},
\author[iggp,ess,nice]{\small{Didier Sornette}\thanksref{EM}}
\address[iggp]{Institute of Geophysics and Planetary Physics, University of California
Los Angeles, CA 90095-1567}
\address[ess]{Department of Earth and Space Sciences, University of California
Los Angeles, CA 90095-1567}
\address[nice]{Laboratoire de Physique de la Mati\`ere Condens\'ee,
CNRS UMR 6622 and Universit\'e de Nice-Sophia Antipolis, 06108
Nice Cedex 2, France}
\thanks[EM]{Corresponding author. Department of Earth and Space
Sciences and Institute of Geophysics and Planetary Physics,
University of California, Los Angeles, CA 90095-1567, USA. Tel:
+1-310-825-2863; Fax: +1-310-206-3051. {\it E-mail address:}\/
sornette@moho.ess.ucla.edu (D. Sornette)\\
http://www.ess.ucla.edu/faculty/sornette/}
\begin{abstract}

In the aftermath of the burst of the ``new economy'' bubble in
2000, the Federal Reserve aggressively reduced short-term rates
yields in less than two years from 6$^{1/2}$ to 1$^{1/4}$~\% in an
attempt to coax forth a stronger recovery of the US economy. But,
there is growing apprehension that this is creating a new bubble
in real estate, as strong housing demand is fuelled by
historically low mortgage rates. Are we going from Charybdis to
Scylla? This question is all the more excruciating at a time when
many other indicators suggest a significant deflationary risk.
Using economic data, Federal Reserve Chairman A. Greenspan and
Governor D.L. Kohn dismissed recently this possibility. Using the
theory of critical phenomena resulting from positive feedbacks in
markets, we confirm this view point for the US but find that
mayhem may be in store for the UK: we unearth the unmistakable
signatures (log-periodicity and power law super-exponential
acceleration) of a strong unsustainable bubble there, which could
burst around the end of the year 2003.

\end{abstract}
\end{frontmatter}

\section{Introduction}

While the US economy has generally been contracting in the last
two years, real estate has been growing: house prices have been
rising at a rate of about 2 per cent a year faster than income
gains. Real consumer outlays and spending on residential
construction each rose about 3 percent during 2001. Meanwhile, the
gross domestic product (GDP) fell about one-half percent, a drop
that would have been far worse without a strong real estate
sector. While stock market losses have destroyed maybe as much as
\$US 5 trillion in investor wealth since the market's peak, there
has been an offsetting effect in the real estate market. Home
equity has gained about \$US 1.7 trillion in the same period,
according to the chief economist at the biggest US mortgage firm,
Fannie Mae. Since, according to the Federal Reserve, home values
have double the impact on consumer spending that stock values have
via the ``richness effect,'' the housing boom has offset almost
two-third of the the stock market losses on the economy.

Such offsets have triggered talks about a real estate bubble in
the US. Investment weekly Barron's claimed to spot a ``bubble
mentality'' last April 2002 and analysts are increasingly
scrutinizing the possible evidence. A managing director of Pacific
Investment Management Co (PIMCO), the largest America bond fund,
agrees there is ``potential for a bubble in the US residential
property market'' as a result of the lowest mortgage rates since
the 1960s. The statistics released every month continue to confirm
that ``the housing sector continues to defy all odds,'' in the
words of the chief economist for the National Association of
Realtors, David Lereah. Sales of existing housing have been and
are continuing to run at a robust if not enthusiastic pace. Total
mortgage debt outstanding has risen sharply during the last
decade. While the total was about \$US 2.7 trillion in the first
quarter of 1990, by the fourth quarter of 1999, it had almost
doubled, to \$US 5.2 trillion. As a comparison, the total amount
of cumulative borrowing by the Federal Treasury, (the national
debt), was about \$US 5.7 trillion in August 2000. American
mortgages are on the path of becoming the single largest class of
fixed income securities on the planet. Add to these elements that
the demand for mortgage borrowing outstrips aggregate domestic
saving (which is currently negative and has reached in the last
months the lowest level since record keeping began in 1959). This
negative saving rate combined with the continuing rapid growth of
mortgage borrowing implies that there must be a reduction in
non-mortgage lending or an increase in fund flows from abroad or
both \cite{Rollmae}. This may lead to an increased instability
through globalization, resulting from the behavior of
international investors \cite{Stiglitz}. To make things look even
worse, the real estate bubble is part of a general huge credit
``bubble'' that has developed steadily over the last decades,
which includes the various US federal money supply, the personal,
municipal, corporate debt and federal debts (estimated by some to
add up to as much as several tens of trillion US dollars), which
may not only drag down the recovery of the economy but also lead
to vulnerability to exogenous crises.

But is there really a real estate bubble?
The science of complexity, which studies
the emergence of organization in systems as diverse
as the human body (biology), the earth (geology) or the cosmos (astrophysics),
suggests novel insights in this troubling question.
The science of complexity explains the spontaneous
occurrence of coherent large-scale collective behavior, such as well-functioning
capitalistic markets but also
financial crashes and depressions, from the repeated nonlinear
interactions between the constituents of economies. This bottom-up mechanism
explains the robustness and strength of modern developed economies as
well as their vulnerability to endogenous instabilities. The theory of
complex systems thus explains the origin of Adam Smith's invisible hand in society
according to which a collection of selfish self-centered
individuals coldly maximizing their individual ``utility functions''
achieve an optimal aggregate social welfare. This theory explains capitalism
and free trade. However, it also explains and predicts the occurrence of
instabilities and of far-from-optimal equilibria, which are inherent in
the bottom-up self-organization \cite{bookcrash}.

Recent academic
research in the field of complex systems suggest that
the economy as well as stock markets self-organize under the competing
influences of positive and negative feedback mechanisms (\cite{bookcrash}
and references therein). Positive feedbacks, i.e.,
self-reinforcement, refer for instance to the fact that, conditioned
on the observation that the market has recently moved
up (respectively down), this makes it more probable to keep it moving up
(respectively down), so that a large cumulative move may ensue.
 ``Positive feedback'' is the opposite of ``negative feedback'', the latter being
a concept well-known for instance in population dynamics in the
presence of scarce resources. Rational markets and stable economic
equilibria derive from the forces of negative feedback. When
positive feedback forces dominate, deviations from equilibrium
lead to crises. Such instabilities can be seen as intrinsic
endogenous progenies of the dynamical organization of the system.
Positive feedbacks lead to collective behavior, such as herding in
buys during the growth of bubbles and in sells during a crash.
This collective behavior does not require the coordination of
people to take the same action but results from the convergence of
selfish interests together with the impact of interactions between
people through the complex network of their acquaintances.

The analysis presented below relies on a general theory of
financial crashes and of stock market instabilities developed in a
series of works. We refer to
Refs.~\cite{bookcrash,SorJoh01,SZ02QF} and references therein for
all details of our approach. In a nutshell, we are looking for
signatures of a faster-than-exponential growth and its decoration
by log-periodic oscillations. The faster-than-exponential
(super-exponential, hyperbolic or power law) growth means that the
growth rate itself grows, signaling an unsustainable regime. We
add the important ingredient that the log-periodic oscillations
have been found to be reliable indicators of endogenous bubbles
signaling a coming instability or change of regime
\cite{Johsorendo}. Using these criteria, we find no evidence
whatsoever of a bubble in the US real estate market. However, the
same analysis applied to the UK real estate market shows that
these two signatures of an unsustainable bubble are unambiguously
present. Since these signatures have been found to be reliable
predictors of past crashes in financial markets, they point to the
end of the UK real estate bubble possible around the end of the
year 2003, with either a crash or a change of regime in the UK
housing market. We should however caution that the crash is only
one possible scenarios according to the theory coupling rational
expectation bubbles with collective herding behavior
\cite{JLS00,JSL99,SZ03prl}.

Technically, we fit the house price indices to the following
versions of the model. In the first version, the logarithm of the
price is given by
\begin{equation}
\ln[p(t)] = A + B(t_c -t)^{m} + C(t_c-t)^{m}
\cos\left[\omega\log(t_c-t)-\phi\right], \label{Eq:lnpt}
\end{equation}
where $\phi$ is a phase constant, $0<m<1$ quantifies the
acceleration of the price, $A= \log[p(t_c)]$ and $B$ and $C$ are
two amplitudes. $B<0$ signals an upward acceleration. This first
version (\ref{Eq:lnpt}) amounts to assume that the potential
correction or crash at the end of the bubble is proportional to
the total price \cite{JS01}. In contrast, the second version
assumes that the potential correction or crash at the end of the
bubble is proportional to the bubble part of the total price, that
is to the total price minus the fundamental price \cite{JS01}.
This gives the following price evolution:
\begin{equation}
p(t) \approx A + B(t_c -t)^{m} + C(t_c-t)^{m}
\cos\left[\omega\log(t_c-t)-\phi\right]~. \label{Eq:pt}
\end{equation}

Finally, we shall use expression (\ref{Eq:fx}) given below, which
incorporates higher-order harmonics beyond the first-order
log-periodic cosine of formula (\ref{Eq:lnpt}). Indeed, the
spectral Lomb analyses reported in section \ref{s2:sign} suggests
the presence of a very strong harmonic at the angular
log-frequency $2\omega$. The possible importance of harmonics was
also noticed and used in our recent analysis of many stock markets
in the anti-bubble regime that started worldwide during the summer
of 2000 \cite{SZ02QF,ZS02}. This followed the analyses of
log-periodicity in hydrodynamic turbulence data \cite{turb1,turb2}
which have demonstrated the important role of higher harmonics in
the detection of log-periodicity. In view of the parsimony and
quality of the fit with expression (\ref{Eq:fx}), we shall use it
for our test of robustness of the estimation of the critical time
$t_c$.

\section{Speculative bubble in UK real estate}
\label{s1:UK}

\subsection{Data sets}
\label{s2:data}

Since 1984, the Halifax house price index has been used
extensively by government departments, the media and businesses as
an authoritative indicator of house price movements in the United
Kingdom. This index is based on the largest sample of housing data
and provides the longest unbroken series of any similar UK index.
We use this monthly house price index data that are retrieved from
the web site of HBOS
plc\footnote{http://www.hbosplc.com/view/housepriceindex/housepriceindex.asp.}.
Six data sets are analyzed as listed in Table~\ref{Tb0}. Each data
set is a time series, which starts in December 1992 and ends in
April 2003.

\subsection{Log-periodic pattern}
\label{s2:LPPS}

In order to identify the speculative real estate bubble in the UK
market, we fit Eq.~(\ref{Eq:lnpt}) to the six data sets described
in Table~\ref{Tb0}, following the fitting procedure described in
detail in \cite{SZ02QF}.

\begin{figure}
\begin{center}
\epsfig{file=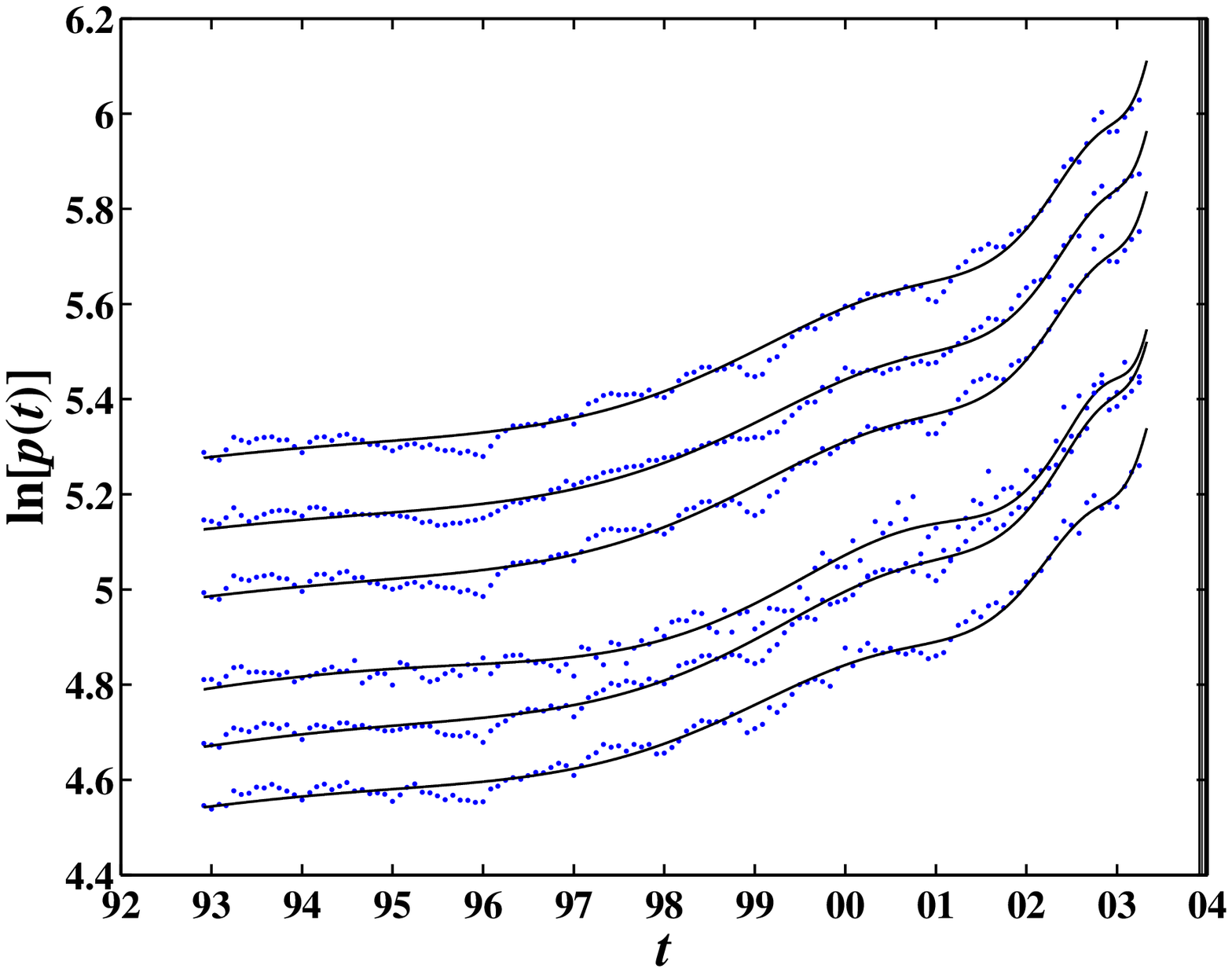,width=15cm, height=12cm}
\end{center}
\caption{Best fits of Eq.~(\ref{Eq:lnpt}) to the logarithm of the
six UK house price indices described in Table~\ref{Tb0} from
December 1992 to April 2003. The values of the fitted parameters
are listed in Table~\ref{Tb1}. The curves have been shifted
vertically downward by $0.15$ incrementally from the first to the
sixth index.} \label{Fig:lnHPI}
\end{figure}

Figure \ref{Fig:lnHPI} shows the best fits of Eq.~(\ref{Eq:lnpt})
to the UK house price index for the six data sets. For the sake of
presentation, we shift the curves downward by multiples of $0.15$:
the curve at the top corresponds to the set 1, the curve
immediately below translated by 0.15 vertically corresponds to set
2, the following curve translated again by 0.15 vertically
corresponds to set 3, and so on. The vertical lines indicate the
values of the critical time $t_c$ obtained from the fit of
Eq.~(\ref{Eq:lnpt}) to each of the six data sets. Recall that
$t_c$ is the expected time for the end of the bubble. It
corresponds to the most probable time for a change of regime or a
crash to occur \cite{JLS00,JSL99}. However, a crash could still
occur before, albeit with small probability.

The log-periodic power-law signatures are clearly visible to the
naked eye. The values of the fitted parameters are listed in
Table~\ref{Tb1}. One observes that the predicted $t_c$ are
consistent for the six data sets and the values of the angular
log-frequencies $\omega$ indicate the existence of a fundamental
angular log-frequency around $\omega \approx 5.8$.

\begin{figure}
\begin{center}
\epsfig{file=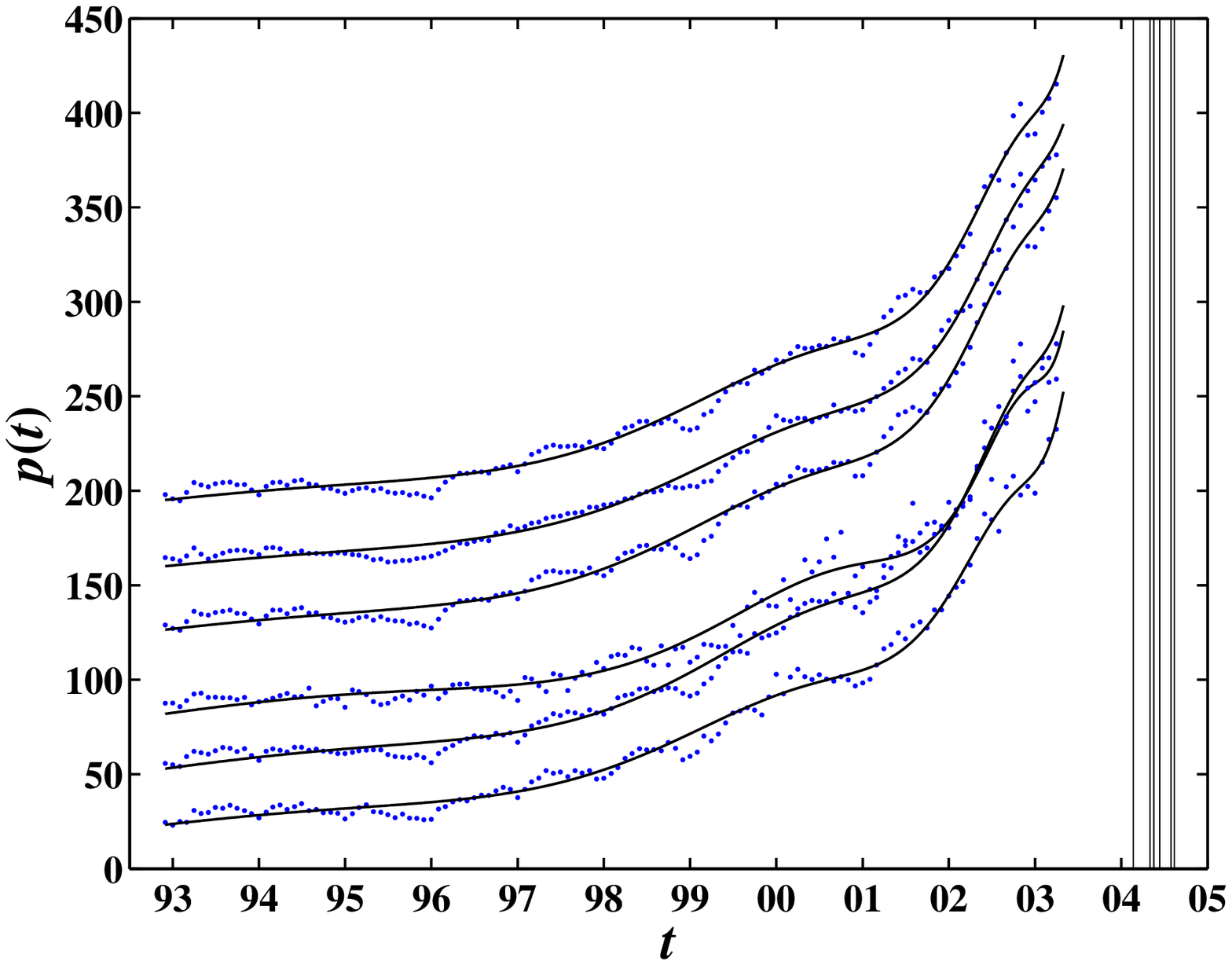,width=15cm, height=12cm}
\end{center}
\caption{Best fits of Eq.~(\ref{Eq:pt}) to the six UK house price
indices described in Table~\ref{Tb0} from December 1992 to April
2003. The values of fitted parameters are listed in
Table~\ref{Tb2}. The curves have been shifted vertically downward
by $35$ incrementally from the first to the sixth index.}
\label{Fig:HPI}
\end{figure}

Figure \ref{Fig:HPI} presents the best fits of Eq.~(\ref{Eq:pt})
to the UK house price index (and not their logarithm as in Figure
\ref{Fig:lnHPI}) for the six data sets described in
Table~\ref{Tb0}. We also translate the curves as in Figure
\ref{Fig:lnHPI} in the same way with a vertical shift of 35
downward for each successive curve. Again, the vertical lines
indicate the positions of the critical times $t_c$ for each data
set. The log-periodic power-law signatures are again clearly
visible. The values of the fitted parameters are given in
Table~\ref{Tb2}. While the predicted critical times $t_c$ are
still quite consistent, they exhibit a significantly larger
scatter than when using Eq.~(\ref{Eq:lnpt}), suggesting that the
bubble is better described by a pure speculative component given
by Eq.~(\ref{Eq:lnpt}). This conclusion is strengthened by the
fact that the fits with Eq.~(\ref{Eq:pt}) give negative exponents
$m$, corresponding to a divergence of the price (in contrast with
the divergence of the change of trend of the price described by
the fits with Eq.~(\ref{Eq:lnpt})). Eq.~(\ref{Eq:lnpt}) seems thus
more consistent with a reasonable price trajectory, exhibiting
super-exponential acceleration, but culminating at a finite value
with infinite unsustainable slope. This again suggests a very
strong component of the bubble component to the price, making the
fundamental component undetectable in this analysis.

In contrast, the found angular log-frequencies $\omega$ are very
robust with respect to the choice of Eq.~(\ref{Eq:lnpt}) versus
Eq.~(\ref{Eq:pt}) and retrieve a fundamental value $\omega \approx
6.8$. This confirms to us the existence of a very clear
speculative signal in these price time series. The consistency of
the results across the six different ways of calculating the House
price indices also suggests that we are deciphering here a very
strong speculative component of the dynamics of house prices,
which remain robust whatever the specific nature of the house and
the way they are estimated.

\subsection{Significance of the log-periodic patterns}
\label{s2:sign}

To assess the significance level of the extracted log-periodic
pattern, we complement the parametric analysis of
Eq.~(\ref{Eq:lnpt}) with non-parametric analyses.

For our first test, we use the first two terms $A + B(t_c -t)^{m}$
of the right-hand-side of equation (\ref{Eq:lnpt}) to detrend
$\ln[p(t)]$, that is, we construct the residual $\{\ln[p(t)]-A -
B(t_c -t)^{m}\}/(t_c-t)^{m}$, in the manner introduced in
Ref.~\cite{JLS00}. A perfect sinusoidal log-periodicity would
correspond to this residual being a perfect cosine
$\cos\left[\omega\log(t_c-t)+\phi\right]$. In order to quantify
the log-periodic oscillatory components of the six time series, we
calculate the Lomb periodograms (which are similar to a Fourier
transform for unevenly spaced data) of the corresponding residues
for the six data sets.

Figure \ref{Fig:Lomb} gives the Lomb periodograms of the detrended
residuals of the six house price indices and their Lomb average as
the thickest line. All periodograms present two significant peaks
around $\omega_1 = 5.9$ and a value $\omega_2 = 13.0$ close to its
harmonics $2\omega_1$.

\begin{figure}
\begin{center}
\epsfig{file=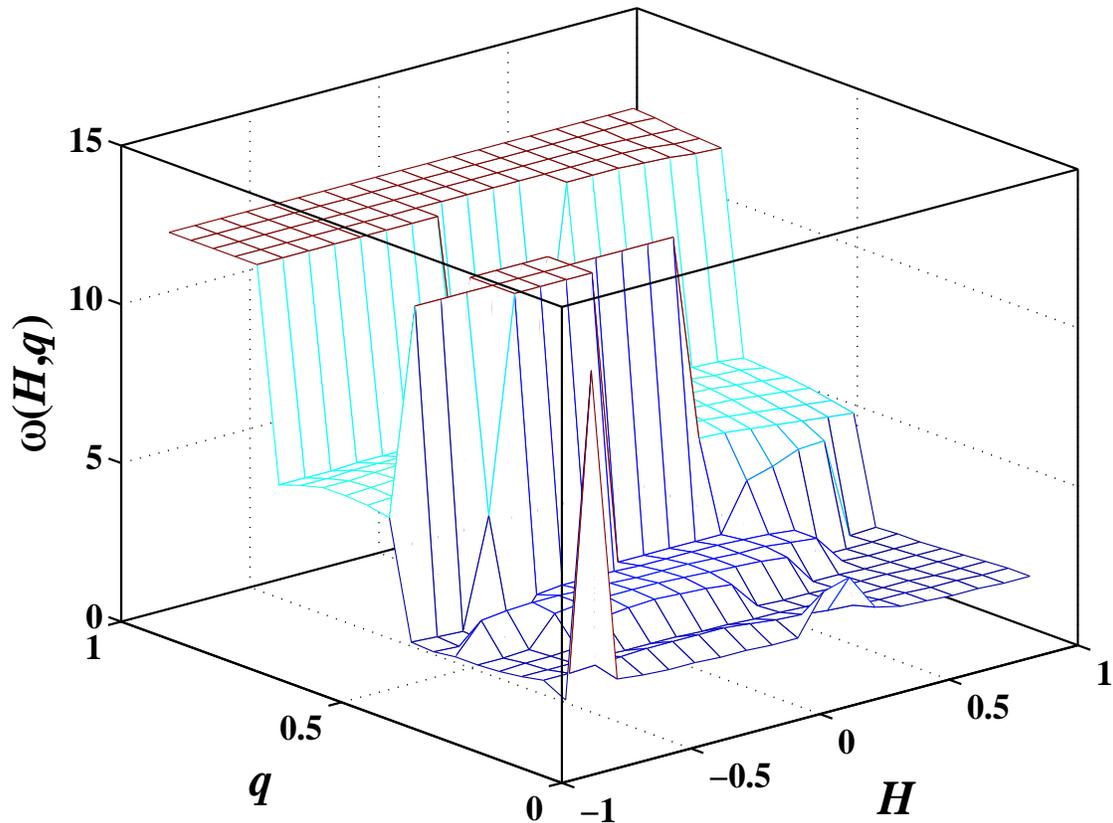, width=15cm, height=12cm}
\end{center}
\caption{Angular log-frequency $\omega$ of the most significant
Lomb peak in each lomb periodogram in the $(H,q)$-analysis of the
logarithm of the price of the seasonally adjusted first time
buyers of all houses (FTB). See text for more explanation. The
middle and high plateaus show respectively the fundamental
log-frequency and its harmonic. The lowest background corresponds
to the artificial most probable log-frequency resulting from the
most probable noise decorating a power law.} \label{Fig:HqAW6}
\end{figure}

Our second test consists in performing non-parametrically the
$(H,q)$-analysis of the logarithm of the price as described in
Refs.~\cite{ZS02PRE,SZ02QF}. In a nutshell, the $(H,q)$-analysis
performs a kind of fractal derivative which is particularly
sensitive to the presence of log-periodicity. The index $q$ refers
to the discrete scaling ratio used in the definition of the
fractal $q$-derivative. The index $H$ is the exponent used to
rescale the $q$-derivative. Scanning $q$ and $H$ provides an
important test of the robustness of the log-periodic signal. A
Lomb periodogram analysis of the $(H,q)$-derivative allows one to
detect the presence of significant log-periodicity. Figure
\ref{Fig:HqAW6} shows the angular log-frequency $\omega$ of the
most significant Lomb peak in each Lomb periodogram for each of
the scanned $(H,q)$ pairs of $(H,q)$-derivative
\cite{ZS02PRE,nonparacrash} of the logarithm of the price of one
of the house price index. The middle and high plateaus show
respectively the fundamental log-frequency and its harmonic. The
lower level corresponds to the artificial log-frequency
corresponding to the most probable noise of a power law signal
demonstrated in Ref.~\cite{HJLSS00}.

The analysis of the power law residuals and the $(H,q)$-analysis
confirm the presence of very significant log-periodicity in the UK
house price indices over the last decade.

\subsection{Parametric fit taking into account the second log-periodic harmonics}
\label{s2:Wei}

\begin{figure}
\begin{center}
\epsfig{file=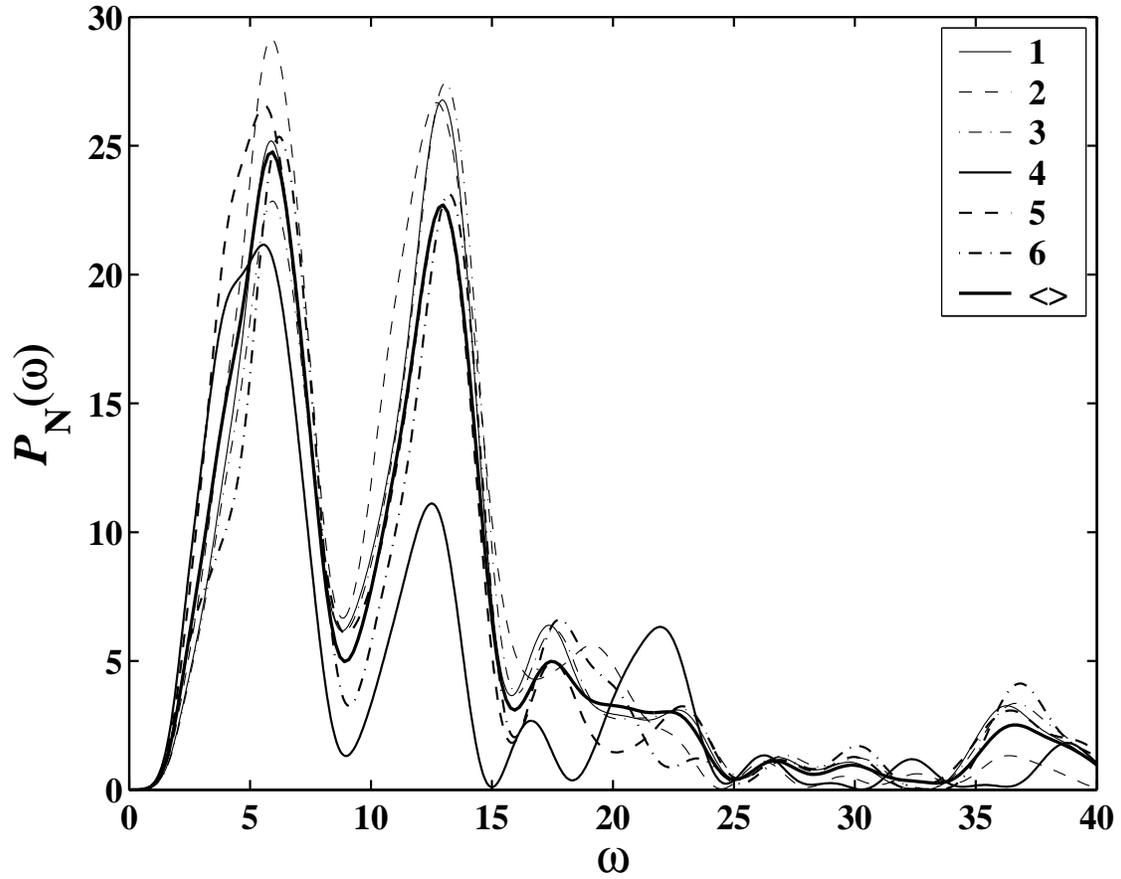, width=15cm, height=12cm}
\end{center}
\caption{Lomb periodograms of the detrended residuals of the six
house price indexes and their Lomb average as the thickest line
(shown in the legend as ``$<>$''). All the periodograms present
two significant peaks around $\omega = 5.9$ and its harmonic
$\omega = 13.0$. } \label{Fig:Lomb}
\end{figure}

Figures \ref{Fig:Lomb} and \ref{Fig:HqAW6} have demonstrated the
strong amplitude of a second log-periodic harmonic. Similarly to
our previous modelling of the stock markets
\cite{SZ02QF,ZS02,ZS03}, we extend (\ref{Eq:lnpt}) into
\begin{equation}
\ln[p(t)] = A + B \tau^{m} + C {\rm{Re}}\left( \sum_{n=1}^N
n^{-m-0.5}{\rm{e}}^{i\psi_n}\tau^{-s_n}\right)~, \label{Eq:fx}
\end{equation}
where $\tau = (t_c-t)/T$ ($T$ determines the time units), $A, B$
and $C$ are three parameters, $N$ is the number of log-periodic
harmonics kept in the description, and
\begin{equation}
s_n = -m+i\frac{2\pi}{\ln\gamma}n~. \label{eq:sn}
\end{equation}
The operator ${\cal \rm Re}(x)$ takes the real part of $x$. In the
second term of the right-hand-side of (\ref{Eq:fx}), we have used
$s_{n=0}=-m$ as seen from (\ref{eq:sn}). Note that the case $N=1$
recovers Eq.~(\ref{Eq:lnpt}). The function (\ref{Eq:fx}) is called
a Weierstrass-type function \cite{GS02PRE}.

Since the choice $\psi_n = 0$ plays an essential role in the
modelling of financial time series in the bubble and/or
anti-bubble regimes as shown in \cite{ZS03,GS02PRE}, we use
Eq.~(\ref{Eq:fx}) with $\psi_n=0$ and follow the procedure
described in \cite{ZS03} to fit the six data sets. This choice,
which corresponds to have the same phases for all harmonics, has
also the advantage of keeping constant the number of degrees of
freedom such that the more complex formula (\ref{Eq:fx}) has the
same number of free parameters as the simplest log-periodic
equation (\ref{Eq:lnpt}). By this model (\ref{Eq:fx}), we thus
gain in relevance while keeping the same parsimony.

\begin{figure}
\begin{center}
\epsfig{file=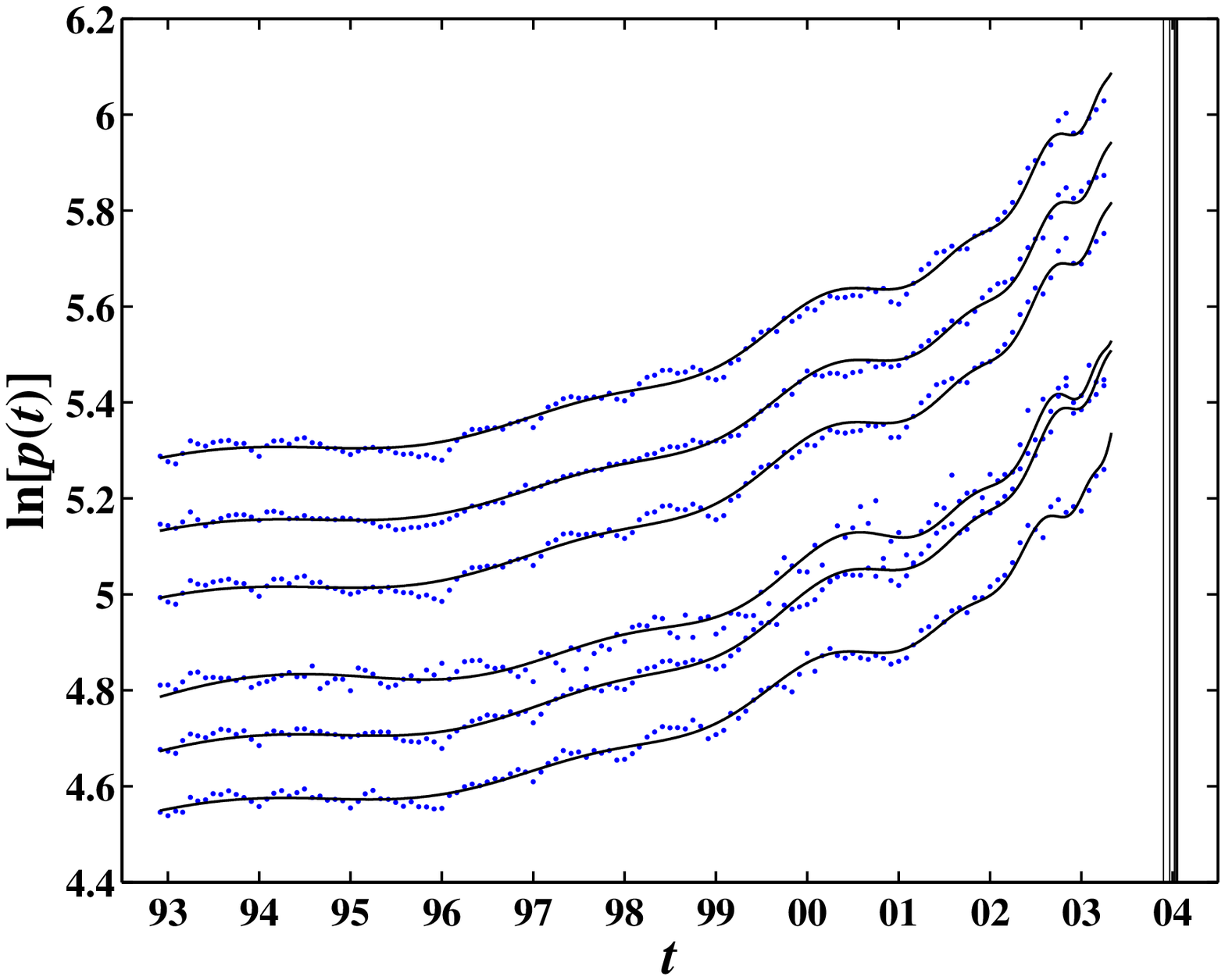,width=15cm, height=12cm}
\end{center}
\caption{Best fits of Eq.~(\ref{Eq:fx}) with $\phi_n=0$ and $N=2$
to the logarithm of the six UK house price indices described in
Table~\ref{Tb0} from December 1992 to April 2003. The values of
fit parameters are listed in Table~\ref{Tb3}. The curves have been
shifted vertically downward by $0.15$ incrementally from the first
to the sixth index.} \label{Fig:Wei}
\end{figure}

Figure \ref{Fig:Wei} shows the best fits of Eq.~(\ref{Eq:fx}) with
$\psi_n=0$ and $N=2$ to the six UK house price indices described
in Table~\ref{Tb0}. The log-periodic signatures are again clearly
visible. The values of the fitted parameters are listed in
Table~\ref{Tb3}. Compared with Table \ref{Tb1}, one can see that
the present fits give much better goodness-of-fit with smaller
r.m.s. errors. Since they have the same number of free parameters,
formula (\ref{Eq:fx}) should be prefered as a better model than
equation (\ref{Eq:lnpt}). The angular log-frequency is found to be
very stable and close to the fundamental value around $6$
identified in previous fits. This increased stability stems from
the fact that expression (\ref{Eq:fx}) automatically accounts for
higher harmonics up to order $N$. This result strengthens the
existence and evidence for strong log-periodicity harmonics. The
exponents $m$ and critical time $t_c$ are consistent across the
six time series. Taking into account higher-order harmonics $N>2$
does not improve the fits significantly due to the coarse monthly
sampling rate and the marginal importance of the third-order and
four-order log-periodic harmonics shown in Figure \ref{Fig:Lomb}.

\subsection{Test of the robustness of the critical time $t_c$}
\label{s2:robust}

The determination of $t_c$ is particularly important as it gives
the estimated termination time of the speculative bubble as well
as the most probable time for a crash, if any. Beyond $t_c$, the
speculative bubble has to transform into another regime because its path
before and up to $t_c$ has become unsustainable. We stress that
several models of stochastic speculative bubbles indicate that a
bubble does not end necessarily with a crash as there is a finite
probability for a bubble to end by a transition to another
regime such as slow deflation
\cite{JLS00,JSL99,hyperbubble}. $t_c$ is thus not the time of the crash,
it is both the end of the bubble and the time at which the crash
is most probable, if it ever occurs. Nevertheless, there is an obvious
interest in estimating when the speculative trend of the real estate bubble in
the UK will end up.

\begin{figure}
\begin{center}
\epsfig{file=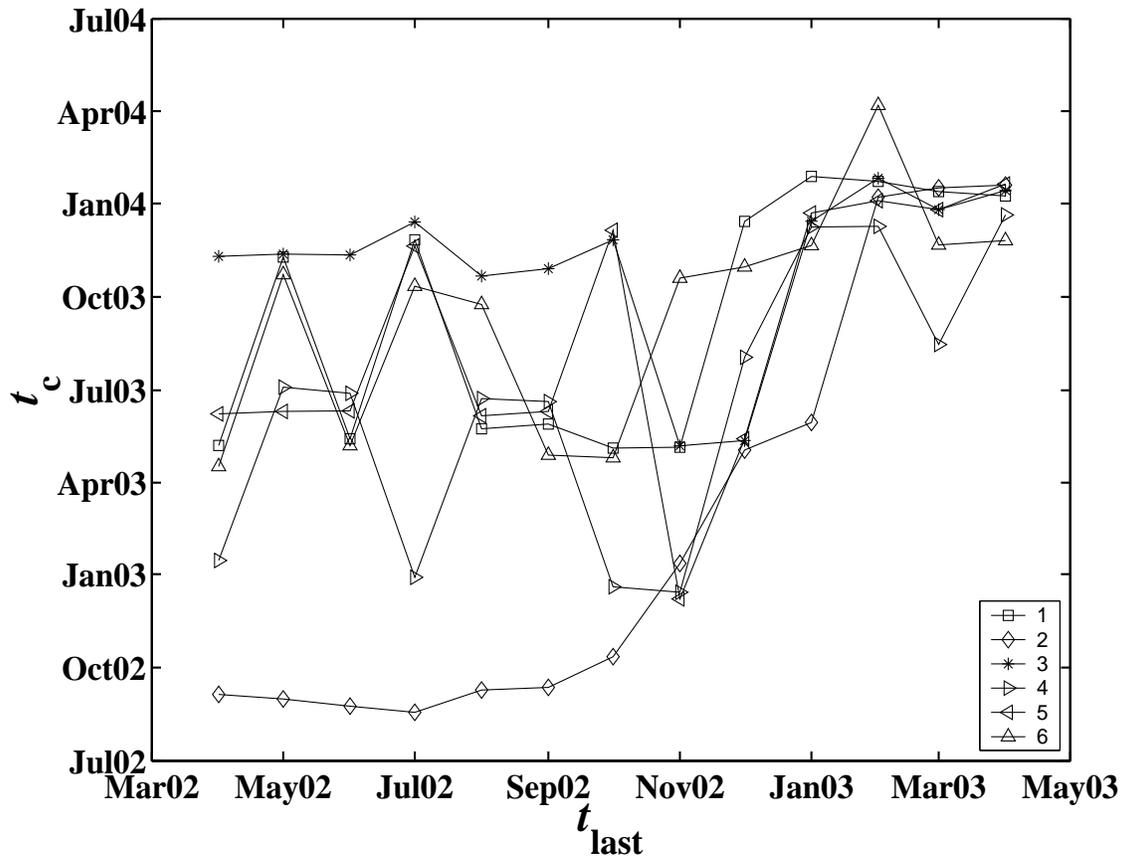,width=15cm, height=12cm}
\end{center}
\caption{Test of robustness of the estimated critical time $t_c$
obtained by varying the last point $t_{\rm{last}}$ of the time
series up to which the fits using the formula
(\ref{Eq:fx}) with $N=2$ and $\psi_n=0$ are performed. See text
for details of the predition procedure.}
\label{Fig:UKRobustWei}
\end{figure}

As an attempt to address this question, we investigate the
robustness of the estimated critical time $t_c$ obtained from the
previous fits. For this, we perform again fits with
Eq.~(\ref{Eq:fx}) of the six time series from December 1992 to a
time $t_{\rm{last}}$ and vary $t_{\rm{last}}$ to see if $t_c$ has
remained approximately constant in the past. For each
$t_{\rm{last}}$, we report in Figure \ref{Fig:UKRobustWei} the
value $t_c$ of the best fit with Eq.~(\ref{Eq:fx}) for each of the
6 time series. Except for time series 2 (all, non seasonally
adjusted), all other 5 seasonally adjusted time series are grossly
consistent with a critical time $t_c$ that could fall anywhere
from the present to the end of the year 2003.

Figure \ref{Fig:UKRobustWei} conveys two pieces of information.
First, the description of the data with the log-periodic formula
(\ref{Eq:fx}) with two log-frequencies is found approximately
stable as a function of time. Second, the precision with which one
can pinpoint the specific end of the presently unfolding
speculative bubble in the UK house prices is limited. In
this vein, the behavior of the critical time $t_c$ predicted with
time series 2 (all, non seasonally adjusted) is instructive: after
a plateau, $t_c$ is found to grow systematically with
$t_{\rm{last}}$. This is a property which has been found in past
investigations (see Chapter 9 of \cite{bookcrash}) to be due to
the fact that the real end of the bubble is beyond the time
horizon over which the log-periodic fits provide reliable
predictors. When using only the first log-periodic harmonics with
expression (\ref{Eq:lnpt}), a monotonous increase of the predicted
$t_c$'s with $t_{\rm{last}}$ is found for all 6 time series, which
is similar to that found with time series 2 using
Eq.~(\ref{Eq:fx}). Together with the results shown in Figure
\ref{Fig:UKRobustWei}, this confirms the stronger relevance of
Eq.~(\ref{Eq:fx}): the addition of a second log-periodic harmonics
seems to extend the time-horizon over which forecasts are stable.
This also shows that the true end of the bubble is beyond the time
horizon of formula (\ref{Eq:lnpt}).

Finally, we should note the significantly smaller variance of the
predicted $t_c$ using for $t_{\rm{last}}$ the last three available data.

\section{Exponential growth in the USA real estate market}
\label{s1:USA}

\begin{figure}
\begin{center}
\epsfig{file=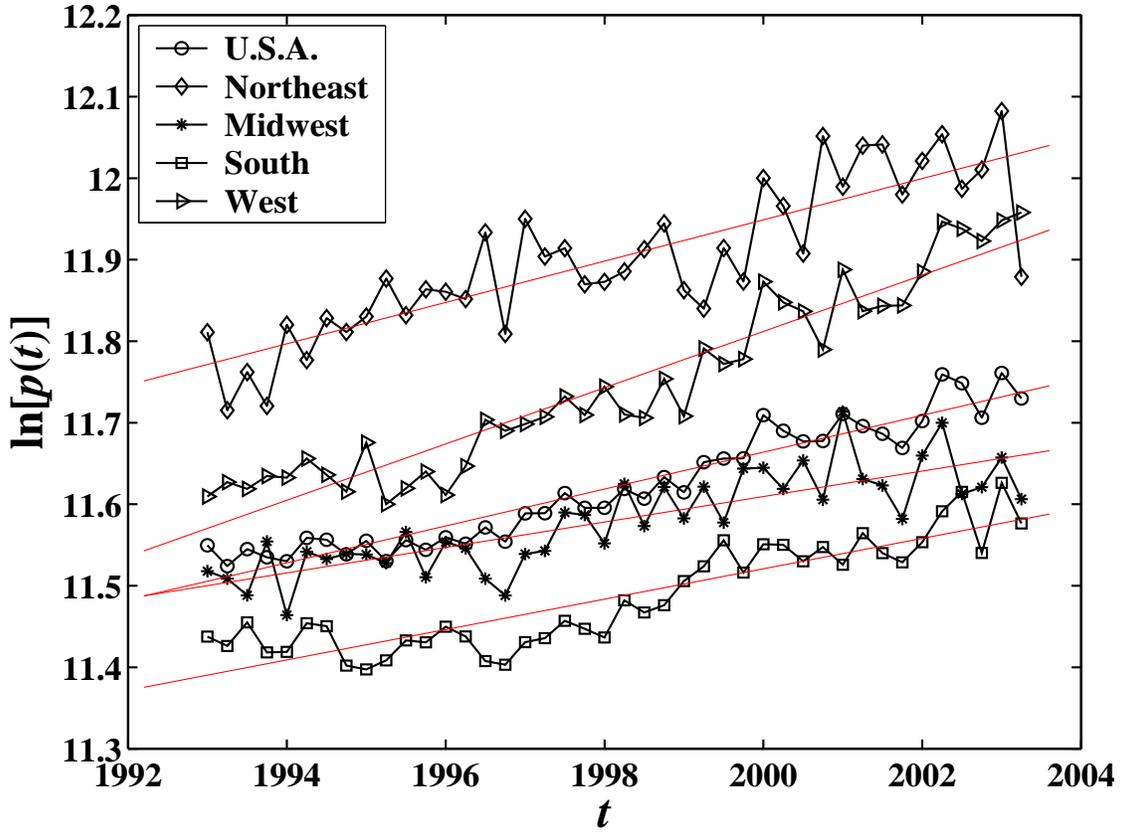,width=15cm, height=12cm}
\end{center}
\caption{Deflated quarterly average sale prices $p(t)$ of
new houses sold in all the states in USA and by regions
(northeast, midwest, south and west) in the last decade as a
function of time $t$ in a semi-log representation. The linear
dependence of $\ln[p(t)]$ against $t$ implies a stable exponential
growth. The yearly growth rates are $2.3\%$ (all states), $2.5\%$
(northeast), $1.6\%$ (midwest), $1.9\%$ (south) and $3.5\%$
(west).} \label{Fig:US}
\end{figure}

Figure \ref{Fig:US} shows the deflated quarterly average sales
prices $p(t)$ of new houses sold in all the states in the USA and
by regions (northeast, midwest, south and west) in the last decade
as a function of time $t$. The data taken from the U.S. Census
Bureau\footnote{http://www.census.gov/const/www/newressalesindex.html.}
are deflated by the consumer price index of the USA available at
the Bureau of Labor Statistics\footnote{http://www.bls.gov/data/}.
The linear dependence of $\ln[p(t)]$ against $t$ in this semi-log
plot qualifies clearly an exponential growth
\begin{equation}
p(t) = p(0) ~{\rm{e}}^{rt}~, \label{Eq:r}
\end{equation}
with a constant growth rate $r$. The yearly growth rates
found by linearly regressing $\ln p(t)$ against time are $2.3\%$
(all states), $2.5\%$ (northeast), $1.6\%$ (midwest), $1.9\%$
(south) and $3.5\%$ (west).
Exactly the same behavior is found (not shown) on non-deflated prices; simply, the
growth rates are different ($4.7\%$
(all states), $4.1\%$ (northeast), $5.3\%$ (midwest), $3.9\%$
(south) and $3.9\%$ (west)).

Using expression (\ref{Eq:lnpt}) to fit these same data sets since
December 1992 to December 2002, we find absolutely no significant
log-periodicity and no power-law criticality in the investigated
price trajectory. According to the proposal
\cite{hyperbubble,bookcrash,Johsorendo} that speculative bubbles
are associated with super-exponential growth with significant
log-periodicity, the recent house price time series in the US do
not qualify as bubbles.

\begin{figure}
\begin{center}
\epsfig{file=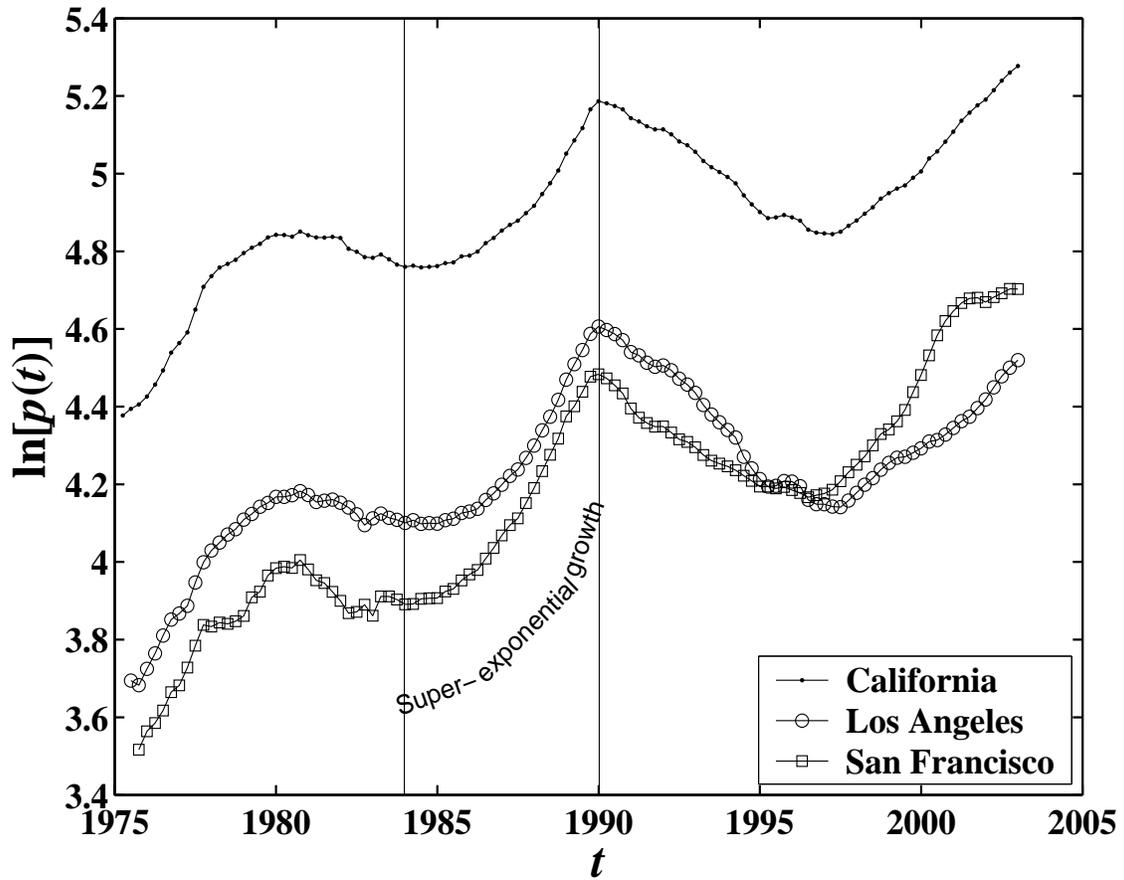,width=15cm, height=12cm}
\end{center}
\caption{The logarithm of the deflated house price indexes
for California, Los Angeles and San Francisco, respectively. The
two vertical lines delimit the super-exponential growth regime
which is characteristic of the real estate bubble that ended in 1990.
} \label{Fig:CA}
\end{figure}

The USA real estate market has been rather segmented because of
its size which is not comparable to the UK market where London
plays a leading role. It would thus be more rational to compare
the UK with California for instance, as the two markets have
approximately the same geographical size and level of integration,
with Los Angeles and San Francisco leading the Californian market.
In Fig.~\ref{Fig:CA}, we plot the logarithm of house price indices
for California, Los Angeles and San Francisco, respectively. The
house price indices are also deflated by the USA consumer price
index for inflation. Note that the data for Los Angeles and San
Francisco are normalized to 100 in the first quarter of 1995,
while that for California is normalized to 100 in the first
quarter of 1980. The difference in normalization dates has no
impact on appreciation rates obtained for each index. The two
vertical lines delimit a super-exponential growth, which ended at
the end of 1989, and which paralleled the Japan real estate bubble
in the 1980's and its burst at the end of 1989. The real estate
bubble ending in 1990 has been well-documented and it is
interesting to find its super-exponential signature so-clearly in
this analysis. For the present, while there is undoubtedly a
strong growth rate, there is no evidence of a super-exponential
growth in the latest six years. Similar conclusions hold for the
un-deflated indices.

\section{Conclusion}

Testifying before the Senate Committee on Banking, Housing and
Urban Affairs in July 2002, Federal Reserve Board Chairman Alan
Greenspan told lawmakers that rising home prices in the USA are a
by-product of ``low mortgage rates, immigration, and shortages of
buildable land in some areas.'' As a result, homeowners have more
equity they can use to pay off high-cost consumer debt and for
other purposes. This leads to a beneficial effect on the US
economy rather than suggesting the possibility of a real estate
crash.

Based on the science of complexity, our analysis provides a
confirmation of this conclusion derived from more standard
economic analysis.

The situation is the opposite for the UK market. Our same analysis
applied to the UK real estate market shows two unambiguous
signatures of an unsustainable bubble, which started years even
before the end of the stock market bubble in 2000. These
signatures have been found to be reliable predictors of past
crashes in financial markets. The analysis points to the end of
the bubble around the end of the year 2003, with either a crash or
a change of direction in the UK housing market. While there are
very strong correlations between stock markets in developed
countries at present \cite{ZS02}, no such correlation has yet
materialized in real estate markets. Investors should however
remain watchful for indications of a possible contagions to the US
in the longer term.

To summarize, what we have learned in this paper is that (i) the
USA real estate market is in a state compatible with  ``rational
expectation'' regime; (ii) the UK real-estate market exhibits an
ultimately unsustainable speculative bubble; (iii) The UK house
prices will continue going up during the year 2003; and (iv) The
Weierstrass-type function (\ref{Eq:fx}) outperforms the simple
log-periodic power law formula.

\textbf{Acknowledgments}

We acknowledge stimulating discussions with D. Darcet and B.
Roehner and a referee for suggestions and thank T. Gilbert for help.
This work was supported in
part by the James S. Mc Donnell Foundation 21st century scientist
award/studying complex system.

\newpage

\begin{table}
\begin{center}
\caption{\label{Tb0} The UK real estate price indices used in this
paper.}
\medskip
\begin{tabular}{cll}
\hline\hline
 No.& Name & Description \\\hline
 1 & All(SA) & All Houses (All Buyers), Seasonally Adjusted \\
 2 & All(NSA) & All Houses (All Buyers), Non Seasonally Adjusted\\
 3 & New & New Houses (All Buyers), Non Seasonally Adjusted\\
 4 & Existing & Existing Houses (All Buyers), Non Seasonally Adjusted \\
 5 & FOO & Former Owner Occupiers (All Houses), Non Seasonally Adjusted\\
 6 & FTB & First Time Buyers (All Houses), Non Seasonally Adjusted\\
\hline\hline
\end{tabular}
\end{center}
\end{table}

\begin{table}
\begin{center}
\caption{\label{Tb1} Parameters of the fits of the log-periodic
function (\ref{Eq:lnpt}) to the logarithm of the real estate price
indices in the United Kingdom.}
\medskip
\begin{tabular}{ccccccccccccc}
\hline\hline
 No.&  $t_c$    & $m$  &$\omega$& $\phi$ &  $A$   &   $B$  &  $C$   &  $\chi$\\\hline
 1 & 2003/12/10 & 0.01 & 5.7 & 3.77 & 35.89 & -28.20 &  0.028 &  0.021 \\
 2 & 2003/12/01 & 0.01 & 5.6 & 5.84 & 35.63 & -27.96 & -0.027 &  0.018 \\
 3 & 2003/12/21 & 0.01 & 5.8 & 1.18 & 37.11 & -29.32 & -0.028 &  0.023 \\
 4 & 2003/11/28 & 0.01 & 5.7 & 0.28 & 33.09 & -25.66 & -0.032 &  0.029 \\
 5 & 2003/12/22 & 0.01 & 5.8 & 4.30 & 37.41 & -29.60 &  0.029 &  0.021 \\
 6 & 2003/12/27 & 0.01 & 6.1 & 0.68 & 34.72 & -27.10 &  0.025 &  0.020 \\
 \hline\hline
\end{tabular}
\end{center}
\end{table}

\clearpage

\begin{table}
\begin{center}
\caption{\label{Tb2} Parameters of the fits of the log-periodic
function (\ref{Eq:pt}) to the real estate price indexes in the
United Kingdom.}
\medskip
\begin{tabular}{ccccccccccccc}
\hline\hline
 No.&  $t_c$    & $m$  &$\omega$ &  $\phi$&   $A$   &   $B$  &  $C$   &  $\chi$\\\hline
 1 & 2004/06/12 & -0.51 & 6.8 & 3.10 & 81.24 & 7953.58 & -309.910 &  5.207 \\
 2 & 2004/08/13 & -0.59 & 6.9 & 1.36 & 98.74 & 13306.96 & 545.197 &  4.351 \\
 3 & 2004/07/29 & -0.56 & 7.0 & 2.15 & 86.32 & 11225.99 & 445.223 &  5.774 \\
 4 & 2004/02/21 & -0.34 & 6.4 & 5.33 & 29.78 & 2573.31 & -95.704 &  7.531 \\
 5 & 2004/05/18 & -0.47 & 6.8 & 5.83 & 59.07 & 6274.64 & 232.723 &  5.713 \\
 6 & 2004/05/02 & -0.44 & 6.9 & 1.00 & 70.50 & 5021.50 & 172.436 &  5.038 \\
 \hline\hline
\end{tabular}
\end{center}
\end{table}

\begin{table}
\begin{center}
\caption{\label{Tb3} Parameters of the fits of the log-periodic
function (\ref{Eq:fx}) to the logarithm of the real estate price
indices in the United Kingdom.}
\medskip
\begin{tabular}{ccccccccccccc}
\hline\hline
 No.&  $t_c$    & $m$  &$\omega$&  $A$   &   $B$  &  $C$   & $\chi$\\\hline
 1 & 2004/01/08 & 0.03 & 6.1 & 17.22 & -9.56 &  0.023 &  0.016\\
 2 & 2004/01/19 & 0.01 & 6.1 & 37.56 & -29.75 &  0.025 &  0.013\\
 3 & 2004/01/13 & 0.03 & 6.1 & 14.97 & -7.28 &  0.022 &  0.018\\
 4 & 2003/12/20 & 0.03 & 6.1 & 16.91 & -9.51 &  0.025 &  0.026\\
 5 & 2004/01/20 & 0.03 & 6.1 & 17.51 & -9.75 &  0.024 &  0.016\\
 6 & 2003/11/25 & 0.07 & 6.1 & 9.49 & -2.29 &  0.015 &  0.016\\
 \hline\hline
\end{tabular}
\end{center}
\end{table}

\clearpage

\end{document}